\documentclass[runningheads]{llncs}
\usepackage[T1]{fontenc}
\usepackage{graphicx}
\usepackage{hyperref}
\usepackage{color}

\usepackage{array}
\usepackage{booktabs} 

\usepackage{xargs} 
\usepackage[pdftex,dvipsnames]{xcolor}  
\usepackage[colorinlistoftodos,prependcaption,textsize=small]{todonotes}
\newcommandx{\unsure}[2][1=]{\todo[linecolor=red,backgroundcolor=red!25,bordercolor=red,#1]{#2}}
\newcommandx{\info}[2][1=]{\todo[linecolor=OliveGreen,backgroundcolor=OliveGreen!25,bordercolor=OliveGreen,#1]{#2}}
\newcommandx{\improvement}[2][1=]{\todo[linecolor=Plum,backgroundcolor=Plum!25,bordercolor=Plum,#1]{#2}}
\newcommandx{\addref}[2][1=]{\todo[linecolor=orange,backgroundcolor=orange!25,bordercolor=orange,#1]{#2}}

\usepackage{amsmath}
\begin{document}
\title{On the Brittleness of CLIP Text Encoders}


\author{Allie Tran\inst{1}\orcidID{000-0002-9597-1832} \and Luca Rossetto\inst{1}\orcidID{0000-0002-5389-9465}}
\authorrunning{A. Tran and L. Rossetto}

\institute{Dublin City University,  Ireland\\ \email{\{allie.tran,luca.rossetto\}@dcu.ie}}

\maketitle              
\begin{abstract}

Multimodal co-embedding models, especially CLIP, have advanced the state of the art in zero-shot classification and multimedia information retrieval in recent years by aligning images and text in a shared representation space.
However, such modals trained on a contrastive alignment can lack stability towards small input perturbations.
Especially when dealing with manually expressed queries, minor variations in the query can cause large differences in the ranking of the best-matching results.
In this paper, we present a systematic analysis of the effect of multiple classes of non-semantic query perturbations in an multimedia information retrieval scenario.
We evaluate a diverse set of lexical, syntactic, and semantic perturbations across multiple CLIP variants using the TRECVID Ad-Hoc Video Search queries and the V3C1 video collection. 
Across models, we find that syntactic and semantic perturbations drive the largest instabilities, while brittleness is concentrated in trivial surface edits such as punctuation and case.  Our results highlight robustness as a critical dimension for evaluating vision-language models beyond benchmark accuracy.

\end{abstract}
\section{Introduction}
\label{sec:introduction}

The rise of large-scale vision-language models (VLMs) has marked a significant advancement in AI, with models like CLIP, BLIP, and Flamingo demonstrating remarkable capabilities on zero-shot tasks \cite{alayrac2022flamingo,li2022blip,radford2021learning,DBLP:conf/cvpr/SinghHGCGRK22,wang2022omnivl,DBLP:conf/cvpr/ZhaiWMSK0B22}. These models are trained on vast datasets of images and text, learning to align visual and textual representations through contrastive learning. CLIP~\cite{radford2021learning}, in particular, combines a vision encoder and a text encoder, allowing it to perform zero-shot image classification by ranking images based on their semantic similarity to a given text query.

Despite their impressive performance on broad benchmarks, a critical, yet underexplored, vulnerability of these models is their surprising sensitivity to minor linguistic variations.
While VLMs are expected to be robust to subtle changes in a text query, small perturbations—such as punctuation, spelling, or word order have been observed to cause large fluctuations in their output \cite{DBLP:journals/access/VadicamoABCGHLLLMMNPRSSSTV24}, severely impacting their reliability in zero-shot tasks.
This phenomenon poses a substantial challenge for their deployment in real-world applications where user inputs are naturally noisy and varied.

To study this phenomenon rigorously, we distinguish between two related concepts. We define \textbf{instability} as the divergence between the image ranking retrieved by an original query and that retrieved by its perturbed version. Instability captures whether the system produces consistent outputs under small input changes.
We then define \textbf{brittleness} as instability \emph{normalized by embedding distance} between the original and perturbed queries.
This measure accounts for the fact that some perturbations cause larger shifts in text embedding space than others: a model is brittle if even a small movement in embedding space results in disproportionately large ranking changes.

This paper provides a systematic analysis of how CLIP’s zero-shot retrieval performance is affected by such perturbations. We study not only semantic changes (e.g., negation, spatial reasoning) but also meaning-preserving variations (e.g., typos, casing, paraphrases). Specifically, we address the following research questions:

\begin{enumerate}
    \item To what extent do small perturbations to the text input affect CLIP’s image rankings (\emph{instability})?
    \item Which types of linguistic perturbations cause the most significant instabilities?
    \item How can we quantify brittleness in a way that distinguishes genuine semantic variation from overreaction to small, meaning-preserving edits?
\end{enumerate}

Our contributions are as follows:
First, we present the first large-scale, controlled evaluation of CLIP’s text encoders under structured lexical, syntactic, and semantic perturbations, spanning 190 TRECVID queries and over one million keyframes.
Second, we introduce instability as a retrieval-oriented measure of robustness, quantified through overlap@$k$ and RBO, and show that instability is systematic but varies across perturbation classes.
Third, we propose a new brittleness index that normalises instability by embedding distance, revealing that CLIP overreacts most strongly to trivial edits (e.g., punctuation, case) rather than to semantic changes.
Fourth, by comparing LAION-trained baselines with fine-tuned (FARE2) and alternative (EVA02-L14) text encoders, we show that architectural and training choices improve robustness but do not eliminate brittleness.
Finally, we discuss the implications of these findings for real-world applications of CLIP, from retrieval to moderation and safety-critical tasks, and sketch directions for mitigation, including brittleness-aware training objectives and query-normalisation layers.
The code used for data processing and analysis is available at \url{https://github.com/allie-tran/clip-brittleness}.

\section{Related Work}
\label{sec:related_work}
\subsection{Prior Work on CLIP's Limitations}

Findings from single-modality research already warn us against over-interpreting deep models as compositional reasoners. In language, shuffling token order degrades BERT much less than one would expect, implying substantial bag-of-words behavior~\cite{hessel_schofield_2021_effective}. In vision, BagNet—a CNN constrained to local features with minimal spatial integration—performs surprisingly well on ImageNet~\cite{DBLP:journals/corr/abs-1904-00760}. This suggests that high accuracy can be achieved by aggregating local cues rather than building holistic scene representations, which in turn leads to confusions about object layout and to collapsing multiple objects into a single description~\cite{newman2024pre,yuksekgonul2023vision}.

Cross-modal evaluations make the weakness explicit. Winoground~\cite{DBLP:conf/cvpr/ThrushJBSWKR22} is purpose-built so that \emph{word order matters}: each example has two images and two captions that use the same words in different orders, with the correct image–caption pairing flipping solely because of order and role binding. CLIP-style VLMs perform only slightly above chance on this setting, indicating poor compositional reasoning even when vocabulary is held constant.
Subsequent studies report sharp drops whenever genuine concept binding is required~\cite{lewis2024doesclipbindconcepts} or when spatial relations must be resolved, as in the What’s Up benchmark~\cite{kamath2023whatsupvisionlanguagemodels}. The visual front-ends used in many multimodal LLMs—often CLIP-based—also exhibit systematic failures on basic visual patterns~\cite{Tong_2024_CVPR}. Beyond composition and relations, core linguistic operators like negation remain unreliable~\cite{alhamoud2025vision,singh2024learn}, and even counting is brittle~\cite{paiss2023teaching}. Taken together, the evidence points to limited structural understanding across modalities rather than isolated edge cases.


\subsection{Why This Brittleness Arises}
Several strands of work trace these failures back to properties of the training signal and the learned embedding geometry. First, contrastive training does not require order-sensitive text supervision: a bag-of-words signal can suffice to train strong zero-shot visual models~\cite{DBLP:journals/corr/abs-2112-13884}. If the loss can be minimized without rewarding composition, models have little incentive to learn it.


Second, analyses of CLIP’s joint embedding space indicate a persistent \emph{modality gap}: text and image embeddings occupy largely disjoint regions with weak alignment~\cite{liang2022mind}. Empirical procedures that reduce this gap yield consistent performance improvements~\cite{liang2022mind}. Subsequent evidence attributes the gap primarily to the contrastive learning objective rather than to dataset mismatch or insufficient training~\cite{fahim2024s}. Moreover, the embedding distribution concentrates on a thin, off-centre ellipsoidal shell, reflecting strong geometric biases~\cite{levi2024double}. These properties may heighten sensitivity to small, meaning-preserving textual perturbations, as minor shifts in the text embedding can traverse decision boundaries even when semantics are unchanged.

Finally, there are emerging theoretical limits. \cite{kang2025clipidealnofix} shows that a single shared latent space cannot simultaneously represent, in a fully faithful way, more than one among basic descriptions, attribute binding, spatial relations, and negation. If so, the failures observed under minor text changes are not just data gaps; they are symptoms of an architectural trade-off built into CLIP’s design.

\subsection{CLIP's Text Encoders}

CLIP’s text encoders instantiate a Transformer architecture \cite{vaswani2017attention}: inputs are tokenised into subwords, embedded, and contextualized via self-attention. Relative to task-tuned encoders such as BERT and RoBERTa, CLIP’s text encoders underperform on language understanding benchmarks like GLUE \cite{devlin2019bert,liu2019roberta,wang2018glue}, while exhibiting strong cross-modal associations between lexical items and visual attributes \cite{chen2023difference}. This contrast indicates an encoder optimised for visual alignment rather than fine-grained linguistic structure. To mitigate sensitivity to textual variation, prior work augments training with paraphrases to stabilize responses to rephrasings \cite{kim2024fine} and proposes defenses against character-level adversarial noise \cite{rocamora2025robustness}. The latter yields gains for various downstream tasks, but the injected symbols and digits are atypical of real user queries. We instead examine linguistically valid, everyday variations, such as orthographic errors, case changes, keyword-style queries, and paraphrases, and assess their impact on \emph{ranking stability} rather than only top-1 retrieval. For deployed retrieval systems, stable rankings under meaning-preserving edits are a primary requirement. Accordingly, we ask: given the geometric and supervisory constraints of CLIP-style training, how volatile are text-to-image rankings under small, realistic changes to the text?

\section{Methodology}
\label{sec:methodology}

We systematically evaluate the brittleness of CLIP's text encoders by applying a series of perturbations to the text input and measuring their impact on the model's output. First, we select a set of CLIP models for our experiments. From a set of queries, we encode the text input using the selected CLIP models and rank keyframes from a video dataset based on their similarity to the text input. The top-ranked keyframes are used as the reference to measure changes introduced by perturbations. We then apply a series of perturbations to the original queries and analyse the changes in the image ranking with respect to the original ranking.

\subsection{Experimental Setup}
We selected a set of CLIP models for our experiments, focusing on those that have been widely used and benchmarked in the literature~\cite{leopold2024divexplore} as well as a fine-tuned variant that has been shown to improve robustness against perturbations~\cite{rocamora2025robustness}. The models are listed in Table~\ref{tab:clip_models}, ranging from the standard CLIP ViT-B/32 to larger models like ViT-H/14. All models are trained on the LAION-2B dataset, which provides two billion image-text pairs, with one model trained on a 400M subset for efficiency~\cite{schuhmann2022laion}. In addition, we include two recently proposed models: FARE2-H14, which is fine-tuned on character-level perturbations to enhance robustness~\cite{abad2025robustness}, and EVA02-L14, which uses a large language model as its text encoder~\cite{wullm2clip}.

\begin{table}[ht]
    \centering
    \caption{CLIP models used in the experiments with their descriptions. Short names are used for brevity in the text.} \label{tab:clip_models}
    \begin{tabular}{p{0.25\textwidth}p{0.72\textwidth}}
        \toprule
        \textbf{Short Name} & \textbf{Description} \\
        \midrule
        LAION5B & \verb|laion5b_s13b_b90k|: Multilingual CLIP ViT-B/32 + XLM-RoBERTa, trained on LAION-5B (13B samples, batch 90k). \\
        LAION-B32 & \verb|ViT-B-32_laion2b_s34b_b79k|: CLIP ViT-B/32, English subset of LAION-5B (2B samples), $\sim$66\% zero-shot ImageNet. \\
        LAION-L14 & \verb|ViT-L-14_laion2b_s32b_b82k|: CLIP ViT-L/14, LAION-2B, balanced accuracy vs. speed. \\
        LAION-L14-400M & \verb|ViT-L-14_laion400m_e32|: ViT-L/14 trained on 400M subset, likely lighter/faster variant. \\
        LAION-H14 & \verb|ViT-H-14_laion2b_s32b_b79k|: Large-capacity CLIP ViT-H/14, trained on LAION-2B, strong benchmark performance. \\
        FARE2-H14 & \verb|ViT-H-14_rho50_k1_constrained_FARE2|: ViT-H/14 LAION-2B fine-tuned on character-level perturbations. \\
        EVA02-L14 & \verb|Meta-Llama-3-8B-Instruct|: LLM2CLIP-EVA02-L-14-336, a large language model being aligned with CLIP's vision encoder. \\
        \bottomrule
    \end{tabular}
\end{table}

We use the queries from the TRECVID~\cite{2024trecvidawad} Ad-Hoc Video Search (AVS) Task from the years 2016 to 2022 for our experiment, resulting in a total of 190 queries, spanning a broad range of visual topics.
Each query contains information about a place, object, action, or any combination thereof.
Since these queries were explicitly designed for the evaluation of retrieval methods, they are inherently suited for our purpose.

We use the first shard of the Vimeo Creative Commons Collection (V3C1)~\cite{DBLP:conf/mmm/RossettoSAB19} as a content collection.
V3C1 contains 100 hours of web-sourced video with a broad and diverse range of content.
It comes pre-segmented into roughly one million video shots, each of which represented by a representative keyframe.
These keyframes are used in our experiments to compare the effects of different query perturbations on the list of retrieved results.

\subsection{Text Perturbations}

We systematically apply various perturbations to the text input to CLIP, categorizing them into three classes:
\begin{itemize}
    \item \textbf{Class 1:} Simple lexical perturbations, such as case changes, typos, deletions, additions, substitutions, and punctuation changes.
    \item \textbf{Class 2:} Syntactic perturbations, such as noun phrase extraction, keyword extraction, and reordering. This class represent how users might naturally simplify or modify queries to focus on key concepts.
    \item \textbf{Class 3:} Semantic perturbations, including paraphrasing and synonym replacement. These perturbations are designed to preserve the overall meaning of the query while altering its surface form.
\end{itemize}

These perturbations are designed to be small enough that they do not fundamentally change the meaning of the query but are sufficient to test the robustness of the text encoders. For example, changing the case of a word or replacing a word with its synonym should not alter the semantic meaning of the query, but it may still affect the ranking of images. Furthermore, to observe the `Bag-of-Words' effect, we also shuffle the order of words in extractive perturbations (e.g. keyword-only, noun-only). We do not shuffle for full-sentence pertubations since the grammatical structure would be lost.

\subsection{Evaluation Metrics}
\label{sec:evaluation-metrics}

We evaluate robustness using two complementary dimensions: (i) the stability of retrieval rankings under perturbations and (ii) the relation between this stability and the magnitude of the perturbation in embedding space.

\paragraph{Text distance.}
First, we measure how much a perturbation changes the text representation itself. For each query, we compute the cosine distance between the original and perturbed text vectors. Larger values indicate stronger movement in embedding space, while smaller values suggest that the perturbation leaves the representation nearly unchanged.

\paragraph{Ranking stability/instability.}
Next, we quantify the divergence in retrieval results. We compare the original and perturbed rankings by computing the Rank-Biased Overlap (RBO)~\cite{webber2010similarity}, which emphasises agreement at the top of the ranking while still accounting for deeper positions. Formally, RBO is defined as:
\begin{equation}
\text{RBO}(p) = (1-p) \sum_{k=1}^{|A|} \frac{p^{k-1}}{k} \left( \frac{|A_k \cap B_k|}{|A_k \cup B_k|} \right),
\end{equation}
where $A_k$ and $B_k$ denote the top-$k$ items in the original and perturbed rankings, respectively.
We define \textbf{instability} as:
\[
\text{Instability} = 1 - \text{RBO}(p=0.99).
\]
This measure captures how much the rankings diverge, with higher values indicating greater instability.

\paragraph{Brittleness.}
Finally, we normalize instability by the corresponding intra- (perturbations within a query) and inter-query text distances to account for the fact that large embedding shifts naturally lead to ranking differences. We define the \textbf{brittleness index} as:
\[
\mathcal{B} = \log \left(\frac{\text{Instability} \times \text{Inter-query Distance}}{\text{Intra-query Distance}}\right).
\]

where \textit{Inter-query Distance} is the average cosine distance between all pairs of original queries in the dataset, which normalises for the overall scale of the embedding space; and \textit{Intra-query Distance} is the cosine distance between the original and perturbed query, which captures the magnitude of the specific perturbation.
This measure highlights overreactions: cases where a small embedding change causes disproportionately large shifts in the ranking. A model with low instability but high brittleness is especially vulnerable to meaning-preserving edits such as typos or case changes.

Taken together, these metrics provide a graded view of robustness. Text distance reflects representational sensitivity, instability captures retrieval-level divergence, and brittleness disentangles disproportionate responses from genuinely large semantic changes.


\section{Results}
\label{sec:results}

We present results in three stages: (i) direct overlap analysis at different cut-offs $k$, (ii) instability distributions and regression confirming differences across models and perturbation classes, and (iii) brittleness, which normalises instability by embedding distance.

\subsection{Overlap Analysis}
\label{sec:overlap}

\begin{figure}[ht]
\centering
\includegraphics[width=1.0\textwidth]{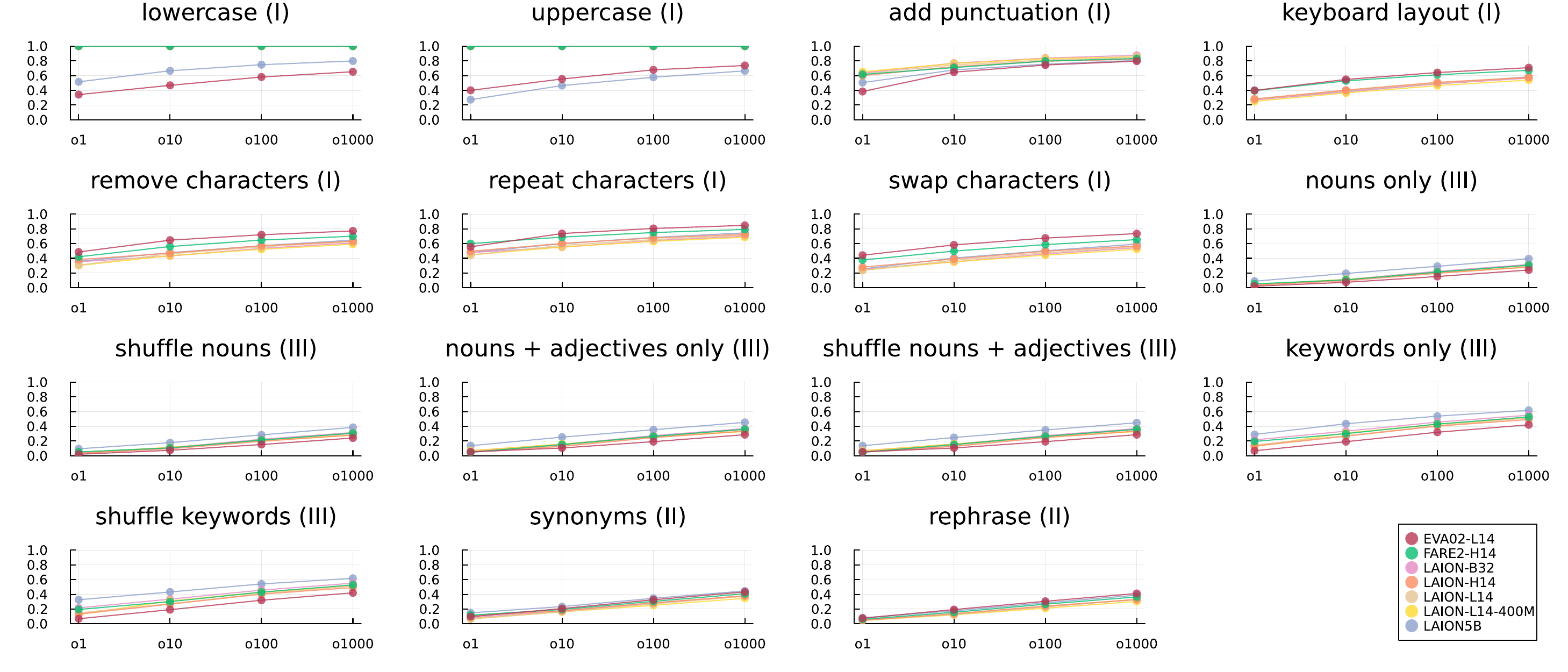}
\caption{Overlap@k between original and perturbed rankings for each perturbation type across models.}
\label{fig:ranking_statistics_overlap}
\end{figure}

Figure~\ref{fig:ranking_statistics_overlap} plots overlap@$k$ between the original and perturbed rankings for each perturbation type. Several patterns emerge.

First, as $k$ increases, overlap rises across all conditions, since agreement is easier to achieve at deeper ranks. However, the critical region for retrieval systems is the head of the ranking (e.g., $k=10$), where differences remain substantial.

Second, most models internally normalise case, leading to identical rankings for lowercase and uppercase variants. More broadly, Class~1 (I) lexical perturbations (case, punctuation, keyboard errors, character swaps) show relatively high overlap. Yet their overlap should arguably be closer to 1.0: for example, adding a single period at the end of a query should not alter retrieval at all. The observed degradation suggests that CLIP's text encoders remain unnecessarily sensitive to surface noise. Among the tested systems, EVA02-L14 and FARE2-H14 show the highest overlap on Class~1 edits, consistent with their training for robustness against such perturbations.

Third, Class~2 (II) syntactic perturbations (keyword-only, noun/adjective-only, and word-order shuffles) considerably lower overlap. Interestingly, shuffling words alone does not produce as dramatic an effect as extracting only nouns or keywords, suggesting that CLIP exhibits bag-of-words behaviour: word order is often ignored, but loss of modifiers or function words destabilises rankings. EVA02-L14 performs consistently poorly in this class, highlighting an ongoing weakness in handling syntactic reduction.

Finally, Class~3 (III) semantic perturbations (synonym substitutions, paraphrasing) show very low overlap, indicating that CLIP’s text encoders are highly sensitive to phrasing and word choice, even when overall meaning remains unchanged. This effect is least pronounced for EVA02-L14, whose LLM-based text encoder is somewhat more tolerant of paraphrastic variation than LAION-based ViTs.

\subsection{Instability Across Models and Classes}
\label{sec:instability}

\begin{figure}[ht]
\centering
\includegraphics[width=\textwidth]{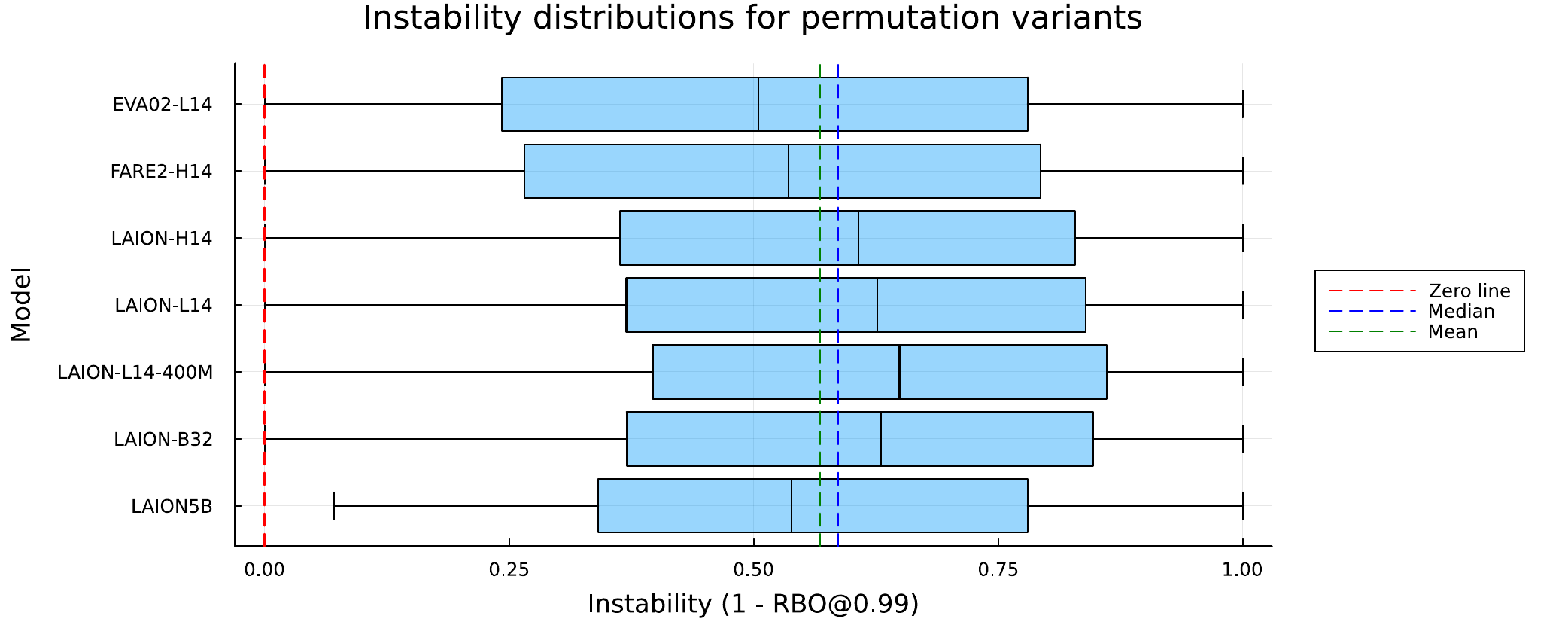}
\caption{Instability distributions across models. Lower values indicate greater robustness. The vertical lines show the average mean and median instability for all models.}
\label{fig:model-instability}
\end{figure}

While overlap curves highlight top-$k$ stability, instability provides a graded summary across the full ranking.
Figure~\ref{fig:model-instability} shows instability distributions aggregated across perturbations: EVA02-L14 and FARE2-H14 are consistently more stable, along with LAION5B, which was trained on multilingual texts.

Mixed-effects regression confirms these trends quantitatively: (i) baseline instability is substantial ($\hat{\mu}=0.568$, $p \ll 0.001$), (ii) model-level coefficients highlight EVA02-L14, FARE2-H14, and LAION5b as significantly more stable, and (iii) perturbation-class effects are strongest for Class~2 and Class~3 variants.

\subsection{Brittleness}
\label{sec:brittleness}
\begin{figure}[ht]
\centering
\includegraphics[width=\textwidth]{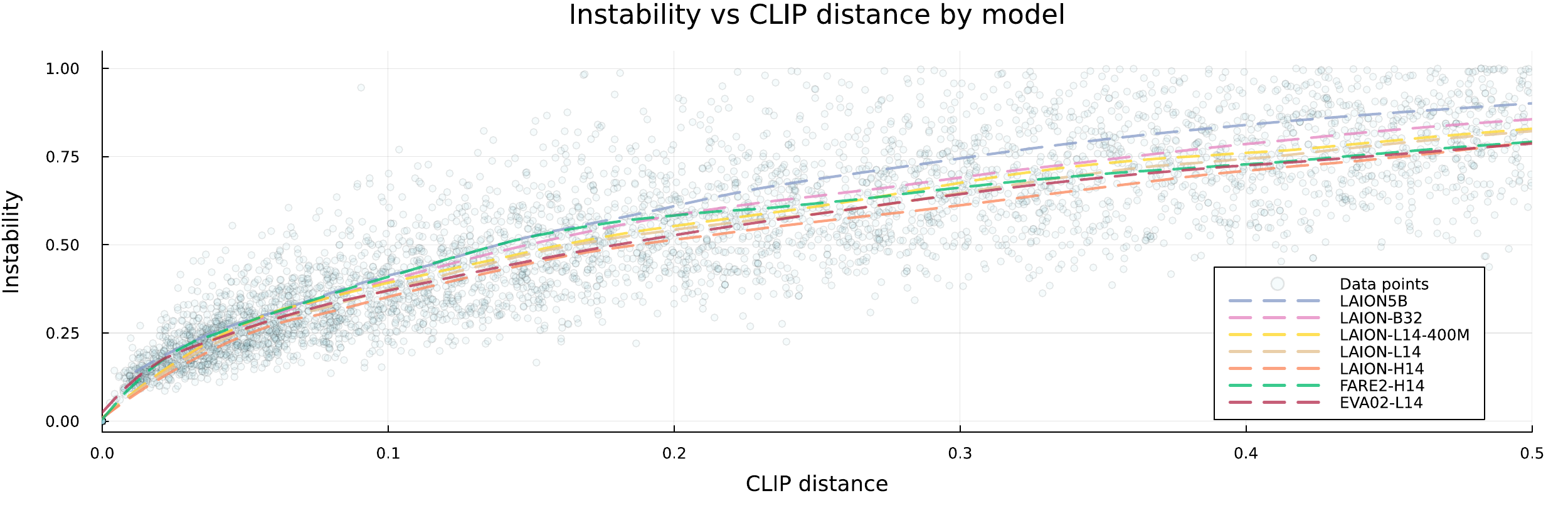}
\caption{Instability vs.\ text distance (normalised by inter-query distance) for a sample of perturbed queries, with LOESS fits per model. Slopes differ across models, indicating varying sensitivity to embedding shifts.}
\label{fig:instability_distance_analysis}
\end{figure}

One explanation is that instability is simply proportional to how far perturbed queries move in embedding space. Indeed, we find a strong positive correlation between instability and text distance (Figure~\ref{fig:instability_distance_analysis}). But the relationship differs across models, and the slope of the regression line varies substantially, indicating that some models are more sensitive to small perturbations than others.

\begin{figure}[ht]
\centering
\includegraphics[width=\textwidth]{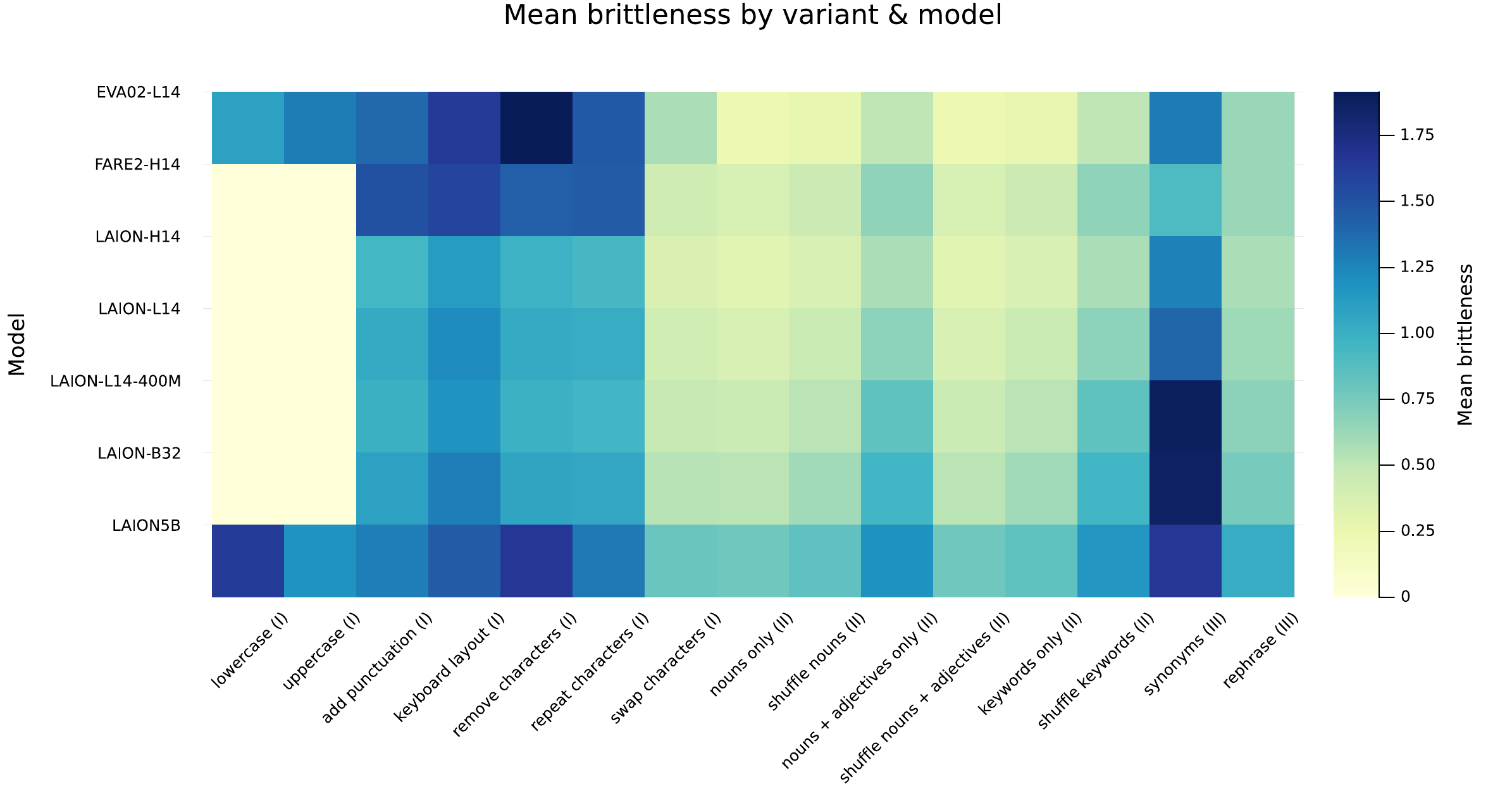}
\caption{Brittleness index across models and perturbation classes. Darker colours indicate higher brittleness. Lexical perturbations (Class~1) stand out as disproportionately brittle.}
\label{fig:brittleness_measure}
\end{figure}

The brittleness heatmap (Figure~\ref{fig:brittleness_measure}) highlights a different pattern than raw instability: Class~1 lexical perturbations induce the highest brittleness, since trivial edits (case, punctuation) cause ranking changes disproportionate to their minimal embedding distance. By contrast, larger embedding shifts from Class~2 syntactic or Class~3 semantic perturbations explain much of their instability, yielding lower brittleness. Depsite having lower instability, LAION5B is remarkably more brittle EVA02-L14 and FARE2-H14 consistently exhibit the lowest brittleness, suggesting that training and architecture refinements mitigate, but do not eliminate, overreaction to surface-level noise.

\section{Discussion}
\label{sec:discussion}

Our analysis reveals three key aspects of CLIP’s behaviour under text perturbations.

\paragraph{Instability is pervasive but structured.}
Across all models, even minor linguistic variations lead to non-trivial ranking changes. The overlap and instability measures show that surface edits (e.g., punctuation, casing) are not entirely ignored, while structural changes (e.g., dropping modifiers, paraphrasing) cause systematic divergence. This confirms earlier warnings that VLMs operate closer to bag-of-words matchers than to genuine compositional reasoners~\cite{hessel_schofield_2021_effective,DBLP:conf/cvpr/ThrushJBSWKR22}.

\paragraph{Brittleness is concentrated in trivial edits.}
Normalising instability by embedding distance highlights that the greatest overreactions occur for formally small changes. Adding a period or altering character case should not affect retrieval, yet these edits yield disproportionate shifts in the ranking. This brittleness points to an undesirable property of the embedding geometry: decision boundaries are so thin that even minimal perturbations can cross them.

\paragraph{Model design matters.}
Comparisons across models show that architecture and training regime shape robustness. EVA02-L14 and FARE2-H14, both trained with additional constraints or alternative text encoders, exhibit consistently lower instability and brittleness. However, neither eliminates sensitivity to paraphrastic or syntactic variation. This suggests that robustness is not simply a matter of scale but also of how alignment objectives are defined.

\paragraph{Implications.}
For downstream applications, instability undermines reliability. In retrieval, unstable top-$k$ lists reduce user trust; in moderation or safety-critical settings, brittleness could lead to inconsistent or unsafe filtering. More broadly, our results imply that evaluation of VLMs should go beyond accuracy on benchmark queries to include measures of robustness under realistic input variation.

\paragraph{Limitations.}
Our study isolates text-side perturbations. While this is appropriate for analysing CLIP’s text encoders, multimodal robustness also depends on visual perturbations and their interaction with text. We evaluated a set of widely used CLIP variants, but further work should include multimodal LLMs that build on CLIP’s encoders. Finally, our brittleness index is one possible normalisation; alternative formulations could capture sensitivity to perturbations differently. Furthermore, we limit the current study to queries in English to include a broader range of CLIP models for analysis. A subset of the analysed models have been trained on multilingual content. Although we have no reason to believe that the presented results are exclusive to English, no analysis of the behaviour of text encoder when faced with non-English queries has been conducted so far.

\paragraph{Outlook.}
The brittleness index suggests potential training interventions: incorporating brittleness penalties into loss functions, or learning projection layers that map noisy queries into more stable embedding regions. Both would encourage models to preserve ranking stability under small, meaning-preserving edits. Ultimately, a robust VLM must treat linguistic variation not as adversarial noise but as an expected feature of human input.



\section{Conclusion}
\label{sec:conclusion}
We have presented a systematic study of the robustness of CLIP’s text encoders under lexical, syntactic, and semantic perturbations. Our analysis introduced two complementary notions: \emph{instability}, the divergence in retrieval rankings caused by input variation, and \emph{brittleness}, which normalises instability by embedding distance to highlight disproportionate overreactions. Using these measures across a large set of queries and models, we showed that instability is pervasive, with syntactic reductions and paraphrastic changes driving the largest divergences, while brittleness is concentrated in trivial edits such as punctuation or casing.

Comparisons across model families revealed that architectural and training refinements, such as FARE2-H14 and EVA02-L14, significantly reduce instability and brittleness but do not eliminate them. This suggests that robustness cannot be taken for granted, even in state-of-the-art vision–language models. For downstream applications, from video retrieval to content moderation, instability undermines reliability and brittleness poses safety risks, making robustness evaluation as important as accuracy benchmarks.

Looking ahead, our brittleness index offers not only an analytic tool but also a potential training signal: it could be incorporated into loss functions to penalise overreaction to meaning-preserving edits. Likewise, projection layers or query-normalisation mechanisms may mitigate brittleness by smoothing the embedding space around trivial variations. More broadly, we hope this work encourages the community to treat robustness to everyday linguistic variation as a core requirement for deploying vision–language models in real-world settings.
\begin{credits}


\end{credits}

\bibliographystyle{splncs04}
\bibliography{bibliography}

\end{document}